\documentclass[tighten, twocolumn]{aastex63}
\usepackage{xcolor, fontawesome, microtype, amsmath}
\definecolor{twitterblue}{RGB}{64,153,255}
\newcommand{\twitter}[1]{\href{https://twitter.com/#1}{\textcolor{twitterblue}{\faTwitter}\,\tt \textcolor{twitterblue}{@#1}}}
\shorttitle{Interstellar communication network. I. Overview and assumptions}
\shortauthors{Michael Hippke}
\begin{document}
\title{Interstellar communication network.\\I. Overview and assumptions}

\author[0000-0002-0794-6339]{Michael Hippke}
\affiliation{Sonneberg Observatory, Sternwartestr. 32, 96515 Sonneberg, Germany \twitter{hippke}}
\affiliation{Visiting Scholar, Breakthrough Listen Group, Berkeley SETI Research Center, Astronomy Department, UC Berkeley}
\email{michael@hippke.org}

\begin{abstract}
It has recently been suggested in this journal by \citet{Benford2019} that ``Lurkers'' in the form of interstellar exploration probes could be present in the solar system. Similarly, extraterrestrial intelligence could send long-lived probes to many other stellar systems, to report back science and surveillance. If probes and planets with technological species exist in more than a handful of systems in our galaxy, it is beneficial to use a coordinated communication scheme. Due to the inverse square law, data rates decrease strongly for direct connections over long distances. The network bandwidth could be increased by orders of magnitude if repeater stations (nodes) are used in an optimized fashion. This introduction to a series of papers makes the assumptions of the communication scheme explicit. Subsequent papers will discuss technical aspects such as transmitters, repeaters, wavelengths, and power levels. The overall purpose is to gain insight into the physical characteristics of an interstellar communication network, allowing us to describe the most likely sizes and locations of nodes and probes.
\end{abstract}


\section{Introduction}
\label{sec:intro}
Humanity started its search for extraterrestrial civilizations (SETI) even before the first crewed space flight \citep{1959Natur.184..844C}. 
We are looking for communications such as optical
\replaced{continuous waves}{pulses}
\citep{2004ApJ...613.1270H,2007AcAau..61...78H,2009AsBio...9..345H}
and \replaced{pulses}{continuous waves} \citep{2015PASP..127..540T} 
or radio signals \citep{1993ApJ...415..218H,2001SPIE.4273..104W,2010AcAau..67.1342S}.
We are observing thousands of exoplanets \citep{2015ARA&A..53..409W}, some of them potentially habitable \citep{2014Sci...344..277Q,2013ApJ...765..131K,2015ApJ...807...45D}. If other advanced civilizations exist, some of them will perhaps be equally curious, and will try to remotely observe interesting targets using telescopes at various wavelengths.

Remote observations of exoplanets with very large telescopes can reveal surface details such as ocean glint, and determine the surface terrains of forests and savannahs \citep{2008Icar..195..927W}. Even multi-pixel imaging is in reach for km-sized apertures, perhaps using the gravitational lens of the sun \citep{2019BAAS...51c..23T}. Remote observations have certain limitations, however. For instance, the direct visualization of meter-sized living organisms is implausible using physical telescopes over parsec distances, because the number of photons from these objects is so small that a receiver must have the size of a Dyson sphere \citep{2010AsBio..10..121S}. Many more geological features can not be examined remotely, but are very interesting to study, such as plate tectonics which might \citep{2017arXiv170706051T,2017PEPI..269...40N} or might not \citep{2014P&SS...98...14N} have a strong influence on habitability. Many habitable (and/or inhabited) planets might be waterworlds \citep{2017MNRAS.468.2803S}, and intelligent life in water and sub-surface is plausible \citep{2019IJAsB..18..112L}, but likely remotely undetectable (unless they produce technosignatures). If humans would have the technology to send a probe into the liquid ocean of the moon Enceladus to search for fish, we would likely try to finance such a mission. Equally, we can argue that other civilizations would be interested in exploring such worlds. The only choice to do so is to use interstellar probes.

Communications will need to take place between interstellar probes, worlds, and migrants \citep{2018IJAsB..17..177W}. Explorative probes will send back images and spectra of exoplanets, and perhaps receive new requests and commands in return. Settlers will signal back greetings to their now distant friends and relatives. Perhaps, at one point, even trade relations will be established, and negotiations for intangible goods, inventions, and art be communicated across the stars \citep{doi:10.1080/14777620801910818,Krugman2010}. Round-trip times likely measure at least in decades to centuries between inhabited exoplanets (or more generally: nodes), making conversations difficult. But useful communication is not necessarily a dialogue. Imagine the internet had been invented three thousand years ago, and would offer a Youtube collection of plays by Sophocles and the documentary about the eruption of Mount Vesuvius near Pompeii. Such an experience would perhaps be preferable to a text-based conversation \citep{ferris1999interstellar}.

Even if humans accomplish none of these feats, others might have done so before us. Other life might have arisen on exoplanets, developed intelligence and interstellar travel, and sent out scouts and probes \citep{2013IJAsB..12...63B}. In a galaxy billions of years old, such things might have happened a long time ago \citep{1976Icar...28..421J,1981Icar...46..293N}, and many times \citep{2018IJAsB..17...96W,2019IJAsB..18..142S}. Probes might be \added{small and smart \citep{tough1998small}},  equipped with artificial intelligence and self-repair capabilities \citep{1974icsp.book..102B}, or posses replication capabilities \citep{1980JBIS...33..251F,1980QJRAS..21..267T} to achieve long lifetimes. \added{It has even been argued that probes could travel and spread between galaxies \citep{2013AcAau..89....1A} over timescales much shorter than a Hubble time \citep{2015CQGra..32u5025O,2016JCAP...04..021O,2017IJAsB..16..176O}, considerably sharpening of the Fermi paradox.}

\begin{figure}
\includegraphics[width=\linewidth]{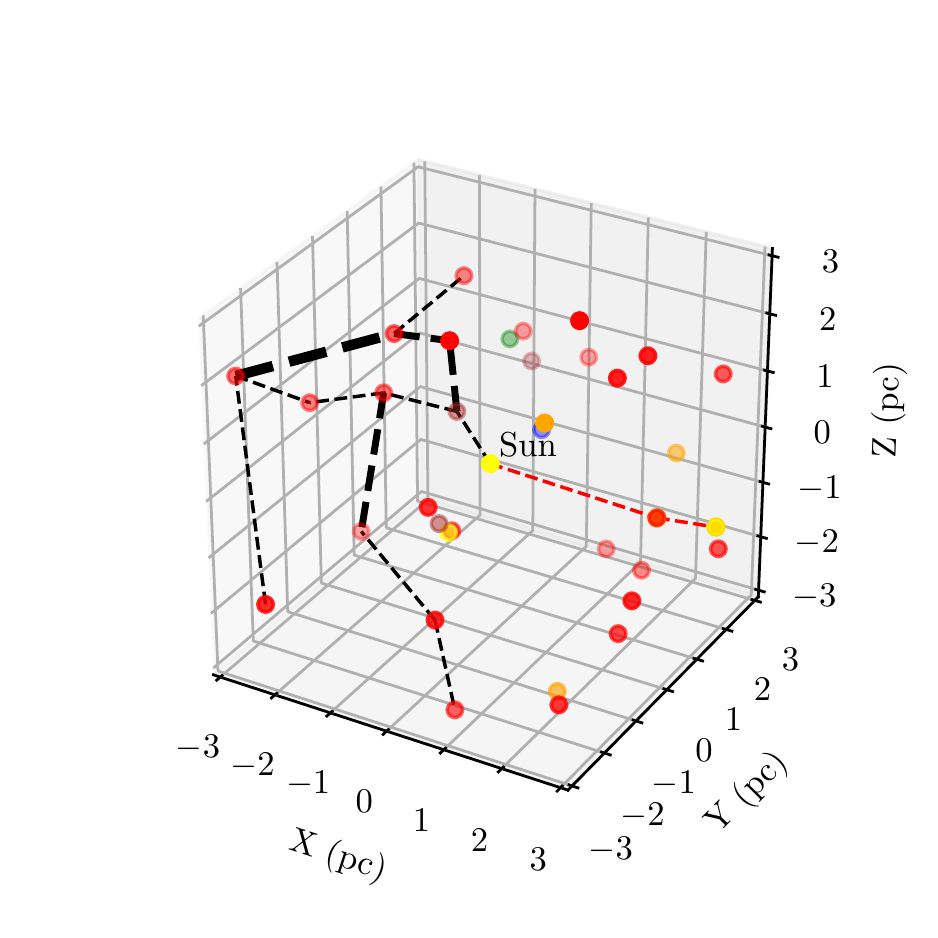}
\caption{Cartesian projection of the 35 closest stars, with the Sun in the center. The dashed red line is the relayed path to Tau Ceti via Gliese 65 \citep{1949AJ.....55...15L}. The dashed black lines show an arbitrary network topology to visualize possible nodes. Thicker lines symbolize higher bandwidths on ``more important'' connections.}
\label{fig:nearest_stars}
\end{figure}

In the literature, interstellar communication is usually considered an end-to-end process (e.g. \citet{1959Natur.184..844C,1961Natur.190..205S,1992MNRAS.257..105B,2015AcAau.107...20M}), although a concept of distributed nodes has been proposed \citep{2015JBIS...68...94H,2016JBIS...69...88G}. The vast distances between the stars pose a challenge for direct connections. Due to the ``terror'' of the inverse square law, a doubling in distance results in a quarter of the data rate, all things being equal. With distances of more than $10^{16}\,$m even between the closest stars, achievable data rates drop quickly over longer distances, even for large apertures and high powers \citep{2019IJAsB..18..267H}. 
In a scenario of more than a few communication partners, it becomes very inefficient to maintain a naive communication structure, where every participant communicates with its partner directly. This becomes very clear in a network with a very large number of clients, such as the internet. Its topology has multiple layers of interconnected domains, with routers in between. While the topology of an interstellar network will be treated in more detail in a future paper of this series, it is clear that end-to-end communications are inefficient for more than a few clients. Simplified, if every participant has one open link, the total network bandwidth $B$ decreases with the square root of the average distance, $B \propto \overline{d}^{-2}$. Such a network scales linearly with the number of participants, but with the square root of its size. Similar scaling relations apply to the Earth's internet \citep{2015arXiv150503449S}. The total (interstellar) network bandwidth can be increased by orders of magnitude if a mesh structure is formed, using repeater station (Figure~\ref{fig:nearest_stars}). The idea has been put forward as an explanation of the \added{(soft version of the)} Fermi Paradox, with the notion that ``nodes might be anywhere'', making it impossible to intercept their tight directed communication beams of laser light \citep{2016JBIS...69...88G,2017JBIS...70..454G}, an issue also raised by \citet{2014JBIS...67..232F}. \added{In other words, machine surrogates could pervade the galaxy \citep{1983QJRAS..24..283B,2018grsi.book.....C}.}

An interstellar network might share some similarities to our own networking, grounded in the common mathematical and physical framework across the universe \citep{russell1989principles}. Networks with very long delay paths and frequent network partitions need to become increasingly delay tolerant \citep{fall2003delay} and make use of forward error correction \citep{baylis2018error}. Nodes will likely operate in a store-and-forward mode and perhaps relay fragmented bundles on multiple paths \citep{2018AcAau.151..401D}. A contact routing graph \citep{araniti2015contact} can be used to prioritize the route with the best chance for earliest expected delivery time. Nodes will need to share meta-information such as their future positions, queue-lengths, and planned availability to compute a best estimate of the optimal routing and transmission time of the bundles. Some of these aspects have been analysed for an interplanetary internet with Mars \citep{2003AcAau..53..365B}.

Consequently, it is worth considering how an interstellar communication network would be built efficiently. With this knowledge, we can start to actually look for it; and if it exists, connect and join. This series of papers will explore the technical aspects of such a communication network. 

\section{Underlying assumptions}
\label{sec:assumptions}
It is useful to state explicitly the underlying assumptions of this communication scheme, and explore the consequences of them being wrong. With increasing knowledge in the future, these foundations can be modified, and changes in the scheme can be made.

\subsection{Other technological life exists}
\label{sec:life}
{\it Assumption A: Humanity is not the only technological species in the galaxy.}

\added{Astrobiology is a young discipline, and the habitability of many locations is not well constrained. Realistically, we do not even know how to precisely define ``habitability''. Asking the question of anthropocentrism, we might wonder whether we inhabit the best of all possible worlds \citep{Leibnitz1710}? From the myriad of habitable worlds that may exist, Earth could well turn out as one that is only marginally habitable. Other ``superhabitable'' worlds could exist \citep{2014AsBio..14...50H}. Planets which are more massive than the Earth might have a higher surface gravity, which would allow them to hold a thicker atmosphere, and thus allow for better shielding of life on the surface against harmful cosmic rays. Stronger surface erosion and a flatter topography could result in an ``archipelago planet'' with shallow oceans ideally suited for biodiversity.

Planets larger than $2\pm0.6\,R_{\oplus}$ are expected to be Neptunian worlds \citep{2017ApJ...834...17C} unsuited for ``life as we know it''. Still, they might have exomoons which can be habitable, and which could offer more total real estate for life than planets \citep{2013AsBio..13...18H,2014OLEB...44..239L,2017A&A...601A..91D}. Life could also develop subsurface \citep{2019IJAsB..18..112L}. More exotic scenarios include biological life in the atmospheres of brown dwarfs \citep{2019ApJ...883..143L}, or in or around binary stars \citep{2016AcAau.128..251V}. 

More conservatively, it has been argued that photosynthesis efficiency on M-dwarf planets is low \citep{2018ApJ...859..171L,2019arXiv190712576L, 2019MNRAS.485.5924L}, and thus the highest probability of complex biospheres could arise for planets around K- and G-type stars \citep{2018JCAP...05..020L}. On the other hand, M-dwarfs are much more numerous and live longer, which could make up for their lower per-star chances of life \citep{1999OLEB...29..405H,2016PhR...663....1S}. In the following, we adopt this conservative and anthropocentric view in order to estimate a cautious lower limit of star systems of interest for interstellar explorers. If the true fraction of interesting systems is larger, then the distance between nodes in a communication network will be smaller.

It has been estimated that of order 10\,\% of stars host ``Earth-sized'' planets within the ``habitable zone'' (with some differing criteria for these requirements) \citep{2013PNAS..11019273P,2018arXiv181108249B,2019MNRAS.487..246Z}, with a range from $2\pm1$\,\% \citep{2014ApJ...795...64F} to $11\pm4$\,\% \citep{2013PNAS..11019273P} for sun-like stars (G- and K-dwarfs) and 20\,\% for M-dwarfs \citep{2015ApJ...807...45D}.}

From the Drake equation, this still leaves the factor $f_i$ (fraction of planets which develop intelligent life) as an unknown. If $f_i=0$, we are the only technological species, and this series of papers can only be taken as an exercise in astro-engineering for humanity. Therefore, we assume $f_i>0$. The factor $f_c$ in the equation (fraction of civilizations that develop a technology that releases detectable signs of their existence into space) is covered by Sections~\ref{sec:exploration} and \ref{sec:travel}, assuming that these signs are actually probes; the longevity aspect $L$ is also discussed in section~\ref{sec:travel}.

With the assumption that there is some non-zero number of space-faring civilizations, we might now wonder about the fraction of stellar system with probes, or the average distance between such system. As an upper limit, we could assume that advanced civilizations have long sent probes to all stellar systems, because the required time to do so is much shorter \citep[$10^6 \dots 10^8$\,yrs,][]{1978JBIS...31..103J,1980QJRAS..21..267T} than the age of our galaxy ($10^{10}$\,yrs). On the other hand, this assumption could be incorrect because many systems might not be worth the effort, e.g. ordinary stars without any planets or peculiarities. While the fraction of stars that have planets is of order unity \citep{2014ApJ...791...10M}, one could argue that sending probes to all ``Hot Jupiters'' is not worth the effort \added{(again, this is a very conservative and anthropocentric perspective)}. Certainly, however, we can argue that exploring planets with intelligent life is sufficiently interesting, and perhaps also all planets with any life, or even all planets which are in principle habitable, e.g. to observe the possible emergence of life under various conditions.

With these assumptions, we can argue for a range of plausible values of $\eta{\rm -interesting}$ ($\eta_{\rm i}$), which is the fraction of stars which warrants sending a probe. The optimistic upper end would be unity \citep{2007IJAsB...6...89B}, with more realistic values of $10^{-1} \dots 10^{-2}$ based on stars with interesting planets in the habitable zone. It is unknown how many of these are truly habitable, but we can adopt an exemplary value of $10^{-4}$, which brings $\eta_{\rm i}$ down to $10^{-6}$, making interesting planets a $6\sigma$ outlier among stars, so that only one in a million stars is worth in situ exploration. This is the lower end of our estimate (although the reader is free to choose other numbers).

From the stellar density of M-, K- and G-dwarfs in the solar neighborhood \citep{1997ESASP1200.....E,2008ApJ...673..864J}, I find the average distance for $\eta_{\rm i}=10^{-6}$ and G-dwarfs as $d_{\rm G}\sim210$\,pc, and $\sim90\,$pc for all MKG-stars. This estimate includes stars of any age, while stars in the inner (thin) disk are generally younger, $0 \dots 8$\,Gyr with a median age of $\approx4$\,Gyr, and stars in the thick disk are older \citep[$8 \dots 12$\,Gyr,][] {2013A&A...560A.109H}. For reference, the total number of stars in the galaxy is $\approx2\times10^{11}$.

This order-of-magnitude estimate holds for the width of the galactic disk \citep[600\,pc,][]{2013A&ARv..21...61R}.
It might differ for other parts of the galaxy which are outside of the ``galactic habitable zone'' (GHZ), argued to range from 4 to 10 kpc from the center of the galaxy (the sun is at 8\,kpc), with 10\,\% of galactic stars in the GHZ, and 75\,\% of these stars in the GHZ are older than the sun \citep{2004Sci...303...59L}. Other estimates see 1\,\% of galactic stars in the GHZ \citep{2011AsBio..11..855G}.

It has been argued that advanced civilizations colonize and inhabit ``clusters'' in space \citep{2016AsBio..16..418L}. The number of clusters would mainly depend on their numbers and longevity. For lifetimes $>1$\,Myr, a single club can be established; for shorter lifetimes, several groups might emerge \citep{2017IJAsB..16..349F}. \added{Numerical studies of galactic habitability also account for colonizing civilizations \citep{2019A&A...625A..98D} and can help to select interesting targets for SETI.}

To conclude, if the fraction of ``interesting'' stars is $1>\eta_{\rm i}>10^{-6}$ (of all spectral types), the average distance between two such stars will be $1 \dots 90$\,pc. This is the plausible distance range of probe-to-probe (or node) interstellar communication. Previous estimates in the literature arrive at similar numbers of $5 \dots 50$\,pc \citep{1975Icar...25..368F}.

\subsection{Exploration}
\label{sec:exploration}
{\it Assumption \replaced{C}{B}: A relevant fraction of ETI wishes to explore the universe.}

Curiosity drives human development, \added{but it is unclear how antropocentric this trait is. Curiosity has pushed humans towards exploration, investigation, and learning about the unknown.} Curiosity is not only the name of a Mars rover \citep{2003JGRE..108.8061C,2013Sci...341.1475G} investigating habitability, climate and geology of the red planet. Curiosity as a behavior is also attributed as the driving force behind human development, and progress in science, language, and industry \citep{Keller1994}. Curiosity is also universal among most animals \citep{Byrne2013}. All breeding life, as we know it, is driven to explore by the inevitable scarcity of resources, filling every ecological niche. Real estate is finite on Earth, and is finite (but very large) in space. It appears reasonable to assume that curiosity is common amongst most or all advanced civilizations.

\added{It appears reasonable to assume that a relevant fraction of ETI is driven towards exploration, perhaps by sentiment similar to our curiosity. It is in fact rational to investigate the unknown; there is a clear evolutionary benefit from exploration, and even more from stealthy exploration. An interstellar probe which secretly observes and relays technology by other species, or more generally: solutions to problems, is very valuable. In contrast, it is quite unclear whether actual communication between species is of positive value. As the ongoing METI debate testifies, contact might even have negative adaptive value, as will be discussed in section~\ref{sec:coord}.}

Thus, there could exist a subset of species which explores, but wishes not to engage with others. This scenario might be described as part of the \textit{zoo}-hypothesis, \citep[``they have set us aside as part of a wilderness area or zoo'',][]{1973Icar...19..347B}. Then, nodes and probes would likely try to hide, and ignore direct contact; making it harder to find them.

\subsection{Possibility of interstellar travel with durable probes}
\label{sec:travel}
{\it Assumption \replaced{B}{C}: Travel velocities are sufficiently high to allow for probes that arrive at their destination intact, with a useful lifetime remaining.}

At the travel velocity of Voyager 2 (17\,km\,s$^{-1}$ relative to the sun) it would take $75{,}000\,$yrs to reach the nearest star, and a billion years to cross the galaxy. Such long times exceed the lifetime of plausible probe designs, and likely the lifetime $L$ of most species \citep{1999JBIS...52...13S,2001SPIE.4273..230R,2019IJAsB..18..445S}. Thus, for our scheme to remain within some boundaries considered sensible today, we require that probes travel between stars on timescales of perhaps $1{,}000\,$yrs or less. As Voyager 2 is still functional after 42\,yrs, including its belt-driven magnetic tape recorder, it appears plausible that advanced technology could allow for probes that last one to two magnitudes longer than that. The required velocity to reach Alpha Cen in a millennium is then half a percent the speed of light. 

A potential technology for such velocities are photon sails with a mass per unit area of order g\,m$^{-2}$. A realistic sail can be accelerated to and decelerated from 0.01\,c solely with stellar photons and magnetic fields  \citep{2017AJ....154..115H,2017ApJ...835L..32H,2018MNRAS.474.3212F}\footnote{This does not apply to stars with low luminosities such as M-dwarfs.}. Other technologies such as fusion-powered spaceships are (perhaps less likely) alternatives \citep{2018haex.bookE.167C}. Without antimatter, space travel velocities are restricted to order 0.1\,c even for very advanced technology if the fuel is taken onboard \citep{1962Sci...137...18V}. Similar velocity limits are expected due to interstellar dust grain collisions \citep{2010AsBio..10..853C,2018JBIS...71..280H}. 

This scheme is agnostic to the question of whether probes carry biological beings, are commanded by AIs \citep{2018arXiv181106526H} or other methods. In any case, it requires capabilities including the calculation of deceleration maneuvers, data collection, re-pointing of transmitters and receivers, and the acceptance of new commands. These requirements appear possible for a wide range of sentience levels onboard the probe.

\begin{table*}
\center
\caption{Transmission times at different data rates
\label{tab:times}}
\begin{tabular}{lccccc}
\tableline
Object & Prefix & Bytes & bit/s (slow) & Mbit/s (medium) & Gbit/s (high) \\
\tableline
Greeting      & KB & $10^3$ & 2\,hrs   & $<1\,$s & $<1\,$s \\
Book          & MB & $10^6$ & 90\,d & 10\,s   & $<1\,$s \\
Video, genome & GB & $10^9$ & \textcolor{orange}{250\,yrs} & 2\,hrs  & 10\,s \\
Holodeck      & TB & $10^{12}$ & \textcolor{red}{$10^5\,$yrs} & 90\,d & 2\,hrs \\
Brain         & PB & $10^{15}$ & \textcolor{red}{$10^8\,$yrs} & \textcolor{orange}{250\,yrs} & 90\,d \\
All human brains & YB & $10^{25}$ & \textcolor{red}{$10^{18}\,$yrs} & \textcolor{red}{$10^{12}\,$yrs} & \textcolor{red}{$10^{9}\,$yrs} \\
\tableline
\end{tabular}
\end{table*}

\subsection{Information valuation}
\label{sec:information}
{\it Assumption D: A relevant fraction of ETI appreciate data of some kind.}

Most work on SETI assume the presence of intelligent beings located on their home planets, distributed over space, with the desire to communicate with each other at the same time \citep[e.g.,][]{1959Natur.184..844C}. In such schemes, low data rates are acceptable, as a simple greeting message would already be considered of the greatest importance.  In this work, we are in principle agnostic to the locations and motivations, but emphasize the value of high data rates. On the most simple level, the reason for such a desire is the existence of a post-industrial society. It is only with the invention of the computer that the processing of vast amounts of data became possible, giving rise to the information age \citep{negroponte2015being}. The digital possibilities currently lead to an economy based on information technology \citep{kluver2010globalization}. In this era, information (i.e., data) are the dominant goods that are produced, valued, and consumed.

The oldest intangible goods in human history are oral traditions, such as songs and stories. In written form, these are today known as articles and books, with data volumes of order MB. On a more advanced level, not words are recorded, but waves: audio and video. Such data can be used to store and exchange a view of historic events (e.g., the moon landing), art, or music. Today, these data are of order GB per piece. These recordings can be pushed towards even more realism in the future, from immersive virtual reality (VR) to full holodeck experiences with 3D recordings of video, audio, smells and other features \citep{2016arXiv160405797M,2018NatRM...3....2H}. Data volumes can be expected at the TB level and beyond. As an example, a dense virtual experience of a cave exhibition on Alpha Centauri could be a valuable trading good.

An even more advanced rationale for a large network bandwidth is a potential post-biological evolution \citep{2003IJAsB...2...65D,2008AcAau..62..499D,2019arXiv191003944G}. For our purpose, it is irrelevant whether this involves sentient artificial general intelligence (AGI), mind-uploading, or similar advances \citep{Kurzweil:2006:SNH:1214678}. It is only crucial that the consciousness (or mind) is treated separately from the substrate (brain, computer), so that it can be digitally transferred to and from different ``hosts'', whatever these are made of. Such a transfer of a mind can be done for any distance given sufficient data rates, ranging from nearby substrates (planetary, interplanetary) to longer distances. Then, interstellar travel can truly be made ``in person'' when the bits which represent a mind are transferred from one interstellar communication node to another. For the individual subjective mind, the travel time is zero; arrival is instant. A trip to Mars takes no longer than one to Andromeda, just the universe has aged a bit more. The curious individual who enjoys traveling has a high valuation for interstellar communication at high data rates\footnote{Purely digital minds might tend to converge to one entity, if many of these are very close to each other in terms of communication proximity. The exchange of thoughts could lead to increased mixing of beliefs, cultures, species. Many minds might sublime into one entity (as perceived by an outside observer). The same might happen to many entities once they travel and interact. Is it one? Is it many? The distinction might become a less useful description. Such a ``group mind'' could be described as an entity which pursues aligned objectives, e.g. builds physical objects, without destructive competition or war.}.

On a related note, the transmission of genomes might be of interest, either to be studied or assembled at a new destination. The human genome has $6\times10^9$ base pairs,

which can be encoded with 1 byte (8 bits) per 4 base pairs in $1.2\times10^{10}$\,bits, or 1\,GB.

\added{Even larger storage capacity is required for} mind-uploading \citep{hauskeller2012my}. The capacity of the human brain is unknown, but for an order of magnitude can be estimated taking its $1.5\times10^{14}$ synapses \citep{pmid12543266} of which each is believed to store on average 4.7\,bits \citep{pmid26618907}, for a total capacity of $7\times10^{14}$\,bits. The efficiency of the human brain is unclear, so that some sort of compression might be possible. On the other hand, evolution should prevent us from extremely inefficient brains (e.g., less than 1\%). A super-mind or AGI might also require a larger data volume, perhaps by a few (e.g., 3) orders of magnitude. This makes for a plausible upper range of data volume of $10^{12} \dots 10^{17}$\,bits. Earth has $10^{10}$ humans with perhaps $10^{15}$\,bits each, for a total of $10^{25}$ bits. One can only speculate about sentient beings with an even larger amount of data.

An overview of these numbers is given in Table~\ref{tab:times}. Data transfer durations $>250\,$yrs are marked in orange, as they approach the timescales of fast interstellar travel; those $>10^5\,$yrs are marked in red, because inscribed matter probes would be the superior solution \citep{2004Natur.431...47R}.

Slow interstellar probes (inscribed matter probes) could contain a lot of data, both in passive storage as well as in active use. Robust storage at the molecular level can achieve $10^{23}\,$bit per gram \citep{2018AcAau.151...32H}. A Matrix-like simulation \citep{grau2005philosophers} of all humans would require $\approx100\,$g in data storage mass, plus about the same amount for shielding and an unknown amount for the actual ship. Small, slow interstellar probes could carry whole populations between the stars, while being essentially undetectable. This is one possible, yet apparently unexplored, form of transcension \citep[or sublimation,][]{banks2008excession,2012AcAau..78...55S} where advanced civilizations appear to leave the universe, yet are still present \citep{2012AcAau..78...55S} as they shrink in ``physical size'' on the Barrow-scale \citep{barrow1999impossibility}. Lastly, minds on a silicon substrate can influence their subjective perception of time by adjusting the clock speed of their computing host \citep{2018arXiv180608561S}. A ``simulation'' running slower will subjectively increase the travel speed, and shorten the time to arrival. In the extreme, hibernation would make arrival instant.

\added{Some minds may decide to hibernate into the distant future to exploit a lower temperature environment useful to increase the number of achievable computations \citep{2016JBIS...69..406S}, although the physics of such a scheme have been challenged \citep{2019FoPh...49..820B}. Hibernation could take place inside black holes \citep{2011arXiv1104.4362V,2011CQGra..28w5015D,2012GrCo...18...65D,2017AmJPh..85...14O}. The possibility of thinking an infinite number of thoughts with finite energy over infinite time has already been raised by \citet{1979RvMP...51..447D}, but similarly challenged \citep{2000ApJ...531...22K}. A threat to such schemes is proton decay, whose existence is equally unclear.}

\begin{table*}
\center
\caption{Recent and speculative future technologies and developments
\label{tab:tech}}
\begin{tabular}{ccccc}
\tableline
Year & Kardashev & Energy & Technology & Space \\
\tableline
1900 & 0.58 & Fossil  & Airplane       &  \\
1970 & 0.67 & Fission & CPU            & Moon landing \\
2020 & 0.72 &         & AI             & Solar system exploration \\
     &      & \textcolor{blue}{Fusion}  & \textcolor{blue}{Strong AI}      & \textcolor{blue}{Mars settlement} \\
     &      &         & \textcolor{blue}{Mind uploading} & \textcolor{blue}{Habitats} \\
     &      &         &                & \textcolor{blue}{Interstellar probe} \\
\tableline
2100 & \textcolor{red}{1.0}  & \textcolor{red}{Antimatter storage}       & \textcolor{red}{Self-replication} & \textcolor{red}{Dyson ring} \\
\tableline
     & 2.0  & \textcolor{red}{Matter-Energy conversion}        &  & \textcolor{red}{Dyson sphere} \\
\tableline
     & 3.0 & \textcolor{red}{Starlifting} &  &\textcolor{red}{Dyson spheres} \\
\tableline
\end{tabular}
\end{table*}

\subsection{Technological baseline}
\added{If the ``doomsday'' is not very near \citep{1993Natur.363..315G}, technology will likely evolve and new inventions will be made. The future ladder of technologies is of course uncertain, but we can again relapse to (hopefully reasonable) assumptions and projections. For example, most people would agree that building a Dyson sphere is much more complex than a settlement on Mars. What is more, a solar system settlement of biological humans is almost certainly feasible within known physics, although it might be too costly or unattractive to actually perform. On the other hand, the construction of a Dyson sphere might (or might not) be so complex that it is not feasible for any ETI in the universe, however advanced in technology. Using similar judgement, we can sketch an outline of technologies which the interstellar communication scheme presented here requires.

As shown in Table~\ref{tab:tech}, Earth has progressed from a \citet{1964SvA.....8..217K} level 0.58 in the year 1900 to about 0.72 today, and the trajectory predicts level 1.0 for the year $\sim2100$, corresponding to $10^{16}\,$W of power. By then, fossil and fission fuel can be expected to be phased out and replaced by fusion reactors and photovoltaics. More advanced energy supplies can be  provided by Dyson spheres, direct matter-to-energy conversion, antimatter storage, or harvesting stars (``starlifting''). It may remain undecided whether such technologies are forever fantasies, as they are not required in our scheme (Table~\ref{tab:tech}, red text). In any case, their use can not be common in nearby galaxies, or we would have seen a myriad of Dyson spheres \citep{2014ApJ...792...27W,2014ApJ...792...26W,2015ApJS..217...25G,2015ApJ...810...23Z}, antimatter propulsion \citep{1986Ap&SS.123..297H}, exhaust gamma rays from relativistic spacecrafts \citep{1991LNP...390..300H}, or high-proper motion objects with infrared excess \citep{2014AcAau.105..547T,1985IAUS..112..505P}.

Over the last century, many new technologies have been invented, such as the microprocessor which allowed for the creation of machine learning and limited artificial intelligence. In the intermediate future, perhaps until the year 2100, we can expect further progress towards general (strong) AI, and related features such as mind uploading. We estimate that, on a similar time scale, the first solar system settlements (perhaps on Mars) can happen, together with other space habitats and the first interstellar probes to the nearest star systems. This is the order of the technological level required to build an interstellar communication network described in this series (Table~\ref{tab:tech}, blue text). Technology beyond this level is outside of the scope of this scheme. Perhaps there is a sort of exponential increase in difficulty for technologies beyond a certain level, somewhere between Kardashev 1.0 and 2.0, which restricts the spread of magic-class machinery.}

\section{Layout of the series}
The approach of this series on interstellar networking is incremental. With the presentation of each aspect (assumptions, receivers, transmitters etc.), feedback and constructive criticism from the community is very welcomed. Through iterative refinement, assumptions can be challenged and boundaries established. Most individual aspects will be discussed very technically, building on the foundation series of interstellar communication, which began with \citet{2019IJAsB..18..267H}. In the following, the currently planned outline of the series is given.

\begin{table*}
\footnotesize
\center
\caption{Location performance matrix for a meter-sized probe \label{tab:probe_location}}
\label{tab:location}
\begin{tabular}{lcccccccccccc}
\tableline
 & LEO & GEO & Moon & L1 & Cruithne &  L2 & Mars & L3 & L4, L5 & KBO & SGL \\
 \tableline
Distance from Earth [au] & $10^{-6}$ & $10^{-4}$ & $10^{-3}$ & $10^{-2}$ & 0.08 &  $10^{-2}$ & 0.37--2.67  & 2 & 1 &  30\dots50 & $>550$ \\
Flux from Earth [W\,m$^{-2}$]  & 240 & 2 &   $10^{-2}$  &  $10^{-3}$ & $10^{-4}$ &   $10^{-3}$ &   $10^{-7}$ & $10^{-8}$ &  $10^{-7}$ &  $\textcolor{red}{10^{-10}}$ &   $\textcolor{red}{<10^{-13}}$ \\
Flux from Sun [W\,m$^{-2}$]    & 1361 & 1361 & 1361 & 1388 &  1361 & 1333 & 593 & 1361 & 1361 &  \textcolor{red}{1} & $\textcolor{red}{<10^{-3}}$ \\
Earth visibility & 1 & 1 & 1 & 1 &  1 & 1 &  0.48\dots1 & $0\dots1$  & 1 &  0.5\dots1  & 1\\
$\alpha$ Cen, SGL visibility  & $\leq0.5$ & $\leq0.5$ & $\leq0.5$  & 1 &  $\leq0.5$ & 1 &  $\leq0.5$  & 1 & 1 &  $\leq1$  & 1\\
Risk of destruction & \textcolor{red}{high} & \textcolor{red}{high} & some & no & no  &  no  & some & no & no & no & no \\
Risk of detection & \textcolor{red}{high} & \textcolor{red}{high} & some & \textcolor{red}{7 probes} & some  &  \textcolor{red}{5 probes} & some & none & low & none & none \\
Earth resolution (optical) [m]  & 0.07 & 12 & 129 & 502 &  $10^3$  & 502 &  $10^{4}$ & $10^{5}$ & $10^{4}$ & \textcolor{red}{$10^{6}$} & \textcolor{red}{$10^{7}$}\\
Earth resolution elements & $10^8$ & $10^6$ & $10^5$ & $10^4$ & $10^3$  &  $10^4$  & 343 & 64 & 127 &  \textcolor{red}{4} & \textcolor{red}{0.23} \\
Gravitationally stable & \textcolor{red}{no} & yes & yes & \textcolor{red}{no} & \textcolor{red}{no} &  \textcolor{red}{no} & yes & \textcolor{red}{no} & yes & yes & yes \\

\tableline
\end{tabular}
\end{table*}

\subsection{Exploration probe}
Exploration probes built by other civilizations are of unknown characteristics in terms of mass and size, and their locations (if they exist) are unknown to us. This part of the series will put constraints on masses, sizes and locations using the assumptions from Section~\ref{sec:assumptions}.

The mass of a spacecraft depends on its use-case. The first human study of an interstellar vehicle was ``Project Orion'' \citep{1965Sci...149..141D, 1968PhT....21j..41D}. It was designed as a ``manned interstellar ark'' for settlers, propelled by a nuclear pulse drive using explosions of atomic bombs behind the craft. The design had a total mass of 8 million tonnes. 

Robotic probes such as ``Project Daedalus'' \citep{1978JBIS...31S...5B} proposed to use a direct fusion drive and required an initial mass of 50,000 tonnes. Even vehicles without onboard propellant, such as Bussard's ramjet to scoop and fuse interstellar hydrogen \citep{Bussard,1969AsAc...15...25F,1975rdss.book.....R,1978JBIS...31..222H}, would be massive. Today, a few decade later, such concepts are mostly considered obsolete. The prime focus of interstellar travel has shifted from colonization and crewed vehicles to lightweight, automatic probes. Freeman Dyson proposed the ``astrochicken'', a self-replicating one-kilogram spacecraft \citep{nygren2015hypothetical}. Taken to the extreme (2020 perspective), probes are planned with masses of order gram, using a light sail for propulsion in combination with a powerful laser \citep{2016AdSpR..58.1093L,2017Natur.542...20P}. If lightweight sail materials and technologies such as graphene or an aluminium lattice can be made into a sail with a mass to surface area of $<$g\,m$^{-2}$, it is possible to accelerate and decelerate from (and to) 0.01\,c solely with stellar photons and magnetic fields \citep{2017AJ....154..115H,2017ApJ...835L..32H,2018MNRAS.474.3212F}. Without the need for reaction mass, a sail is a good long-term solution for ``Lurkers'' \citep{1963P&SS...11..485S}.

The minimum mass and size of such sailing probes can be explored with an assumption of the payload mass, and the mass to surface area of the sail. Even for extreme miniaturization, the payload mass has a minimum due to its requirements for shielding against dust and cosmic ray impacts \citep{2017ApJ...837....5H,2018AcAau.151...32H}. Microscopically small (sub-)probes might be useful to explore inside the atmosphere of a planet, and could relay back data to a larger probe in space. An upper limit comes from the fact that apparently no alien megastructures are present in the (inner) solar systems, such as O'Neill cylinders \citep{1974PhT....27i..32O} or World Ships \citep{2012JBIS...65..119H}.

Regarding the locations of ``Lurkers'', it appears useful to make explicit the possible and assumed requirements. As a first draft, consider the metrics listed in Table~\ref{tab:location} for various locations. One may plausibly hypothesize that the purpose of a probe's presence is the gathering of information about planet Earth, including observations in the optical spectrum. After all, most flux from Earth is emitted between 0.3 and $1\,\mu$m. Accepting the laws of known physics, the amount of information which can be gathered is a function of distance, aperture and wavelength (for which we may take $\lambda=0.5\,\mu$m). It appears equally preferable to resolve Earth into multiple elements (pixels); unresolved observations would be possible from afar. Taking a finite aperture (which is not absurdly large), the distance between the probe and the Earth gives the number of resolution elements. For example, Earth can be resolved into a grid of $350\times350\,$pixels with a 1\,m aperture from Mars, but is effectively unresolved from the Kuiper Belt. 

A sail as the method of propulsion makes landing on a surface difficult or even impossible without requiring extremely advanced technology. Furthermore, why land a sailcraft on the rock surface of a co-orbital object like Cruithne? A sail capable of autonomous deceleration could equally set a sailing course {\it near} an asteroid, or essentially anywhere. We will explore this aspect further during the series, and will propose a previously unexplored location for such a probe, optimized under our assumptions.

\subsection{Deep space nodes}
As described in Section~\ref{sec:intro}, network bandwidth can be increased by orders of magnitude when using relay nodes, due to the inverse square law. With our assumption~\ref{sec:travel} of interstellar travel being possible, perhaps via lightweight solar sails, we are also free in our choice of where to place the relay stations. 

A first possibility would be that exploration probes are acting double duty as observers and relays. Indeed, in some scenarios this may be a good choice. For example, a probe that is not very busy can allocate a useful fraction of its data throughput to relay data, in addition to transmitting its own observations. On the other hand, if the construction of interstellar probes is possible, then the construction of two probes per system is only (at most) twice as expensive. An advantage of having a second probe per system is that it can be placed further out from the star, near the gravitational lens. Such a placement allows for gains of order $10^9$ compared to the direct path, and increases data rates by order of $10^6$ \citep{2018AcAau.142...64H}.

Nodes may vary in their physical characteristics to offer bandwidth as needed. Similar to the internet, there might be ``information highways'' as well as lower bandwidth paths. Some of these characteristics may be dictated by the participant's needs, such as an increased wish to communicate more information between certain locations. In some cases, the environment may encourage a certain routing. For example, the lens gain is a function of stellar mass, which makes the smaller M-dwarfs unattractive as major hubs.

To achieve a similar data rate without gravitational lensing, one could place $n$ equally spaced classical nodes in the void between the stars. Given advanced technology, they could be powered by matter to energy conversion, and relay data from each node to the next. The problem with this approach is that the lensing gain is so large that it requires very many nodes to match it. A rough calculation indicates $n \gtrsim 10^4$ for each one node using lensing. Equivalently, one would need to place a node every few light-hours in empty space. A detailed analysis will be given in a future part of this series.

Recent theoretical advances in the description of the wave-optical properties of the gravitational lens now allow for a precise description of its properties, including its point-spread function and the characteristics of the Einstein ring \citep{2017PhRvD..96b4008T,2019PhRvD..99b4044T}. These properties can be used to describe the minimum and optimal physical size of a receiver in the focal plane, and its exact position in the heliocentric reference frame.

Given the position in the solar system reference frame, we can determine the node's position on the sky as a function of transmitter location. There are five bodies to consider: The transmitter near a distant star, the distant star, the sun, an observer on Earth, and the probe in the lens plane. All bodies are in constant motion. We will calculate the locations with respect to each other, and examine the changes over time and associated uncertainties. As we will see, positions can be determined to arcsecond accuracy, making targeted observations (and/or messaging) possible. Active signalling could be initiated using sufficiently powered laser pulses ($\geq$kJ) with ground-based meter-class telescopes.

\subsection{Network bandwidth}
The argument that nodes increase network bandwidth by orders of magnitude must be examined in detail. While intuitively clear that repeaters increase data rates, everything else is unknown. How much faster does data flow? How many nodes should be used? How far should they be set apart? We will describe a range of cost functions to build, transport and maintain stations, plus utility functions (valuations) of bandwidth, latency and reliability. In this network scheme, the number of nodes and their placements will be simulated. This opens the question of the technical properties of a galactic internet: Its network topology, coverage, and handshake protocols.

\subsection{Probe lifetimes}
With some insight into the range of cost functions to build, transport and maintain stations, we should make explicit the issue of finite lifetimes. If self-replication is not feasible (or not cost-efficient), repair will be imperfect, and at some point a probe will need replacement. However, even fast probes (order 10\,\% the speed of light as proposed by ``Starshot'') will have travel times of decades to centuries as the distances to the stars are so large. More robust probes will have a higher probability of arriving at the target intact, but will cost more to build. The law of diminishing returns predicts that the last few percent in durability are the most expensive, and at some point there is a trade-off where it is more cost efficient to produce more probes instead of increasing their longevity, and re-send those that fail. We explore these timescales using a wide range of parameters for probe velocities and travel distances. The failure rate over time is modeled with a multi-parameter Weibull distribution for the different stages in a probes' life: Acceleration, hibernation, deceleration, and exploration.

\subsection{Optimal encoding}
We will examine possible encodings in an interstellar network. The capacity of communication through bosonic channels is bounded to the \citet{holevo1973bounds} limit. This limit can be achieved in the lossless and noise-free case using an ideal photon counter \citep{1993PhRvL..70..363Y,1994RvMP...66..481C} as detailed by \citet{2004PhRvL..92b7902G}. For an introduction, see \citet{2017arXiv171205682H}. Current technology, however, can not achieve $\Delta t \Delta f\sim1$ beyond radio because femtosecond photon counters do not yet exist.

With losses and noise, the upper limit is known \citep{2014NaPho...8..796G}, but very difficult to reach \citep{2003PhRvL..90p7906F}. There appear to be distinct regimes of optimal receivers where homodyne and heterodyne detection approach optimality \citep{2014PhRvA..89d2309T}. Popular modulation techniques in the interstellar communication literature propose channel coding \citep{2013arXiv1305.4684M}, block location coding \citep{2015AcAau.107...20M}, and bandpass filtering with pulse-position modulation (PPM) \citep{2018arXiv180107778L}. All of these are only competitive in certain corner cases. For instance, PPM approaches optimality for a very low number of modes, for instance in very narrow (spectral) bandwidths. These limitations are usually not made explicit, and thus the proposed method is often not preferable.

On the other end of the spectrum, it has been claimed that sufficiently advanced communication technology is entirely ``indistinguishable from noise'' \citep{2004AmJPh..72.1290L}. The argument is that the most information-efficient format for a given message is in the form of a black body spectrum, which maximizes the entropy of photons by offering the largest possible number of encoding modes of bits per photon in a communication channel \citep{1990IJMPC...1..355B,1993PhRvL..70..363Y,2004PhRvA..69e2310G,2017JSP...169..374F}. However, the optimal spectrum is only a one-dimensional blackbody\footnote{If the transmitter is spatially resolved by the receiver ($D(f)>1$), the maximum number of transverse modes allowed by diffraction can be used. Then, the optimal spectrum is a 3D blackbody. For this case, our current knowledge of the optimal spectrum is incomplete.} for the special case where all of the power transmitted is intercepted by the receiver, but only one spatial mode is used (Fresnel number $D(f)=1$). Such an emitter can be distinguished from natural sources because \replaced{all}{most} known \deleted{luminous} {\it astrophysical} bodies are {\it not} perfect blackbodies, with \deleted{the only} exceptions being the Schwarzschild radiation of a black hole \citep{1975CMaPh..43..199H} \added{, the cosmic microwave background \citep{2006MPLA...21.1495H}}, and DB white dwarfs \citep{2017arXiv171101122S,2019A&A...623A.177S}). \added{An artificial construct which processes information at maximum efficiency would be very close to the CMB temperature and could be incidentally hard to detect \citep{2006NewA...11..628C}.}

In the unresolved far-field case ($D(f)<1$), the optimal spectrum is not a blackbody, but a monotonically increasing function up to the maximum frequency $f_{\rm max}$ with a lower cap at $f_{\min} \sim 0.1f_{\rm max}$. Some practical issues are yet unknown, e.g. the spectral effect of variable losses as a function of wavelength. This case is most applicable for our interstellar mission example of a small optical transmitter. It is assumed (and under active research) that the effect of such limitations result in an optimal spectrum which is still a blackbody, but exponentially suppressed and with a higher temperature.

In this paper of the series, I will strive to describe the relevant regimes and techniques.

\section{Discussion}

\subsection{The importance of making assumptions}
The postulation of alien probes inside the solar system is interesting, but more is required to narrow down the most likely locations. Generic ``Lurkers'' have been proposed to be located 
on nearby co-orbital objects \citep{1995Obs...115...78S,1998Obs...118..226S,Benford2019} for surveillance,
near the Earth-moon Lagrangian points \citep{1980Icar...42..442F,1983Icar...53..453V},
in geocentric, selenocentric, Earth-Moon libration, and Earth-Moon halo orbits \citep{1983Icar...55..337F},
in a tight orbit around the Sun \citep{2018JBIS...71..375G},
in the asteroid belt \citep{1978QJRAS..19..277P,2011IJAsB..10..307F},
the Kuiper belt \citep{2012AsBio..12..290L},
on the surface of the moon \citep{2013AcAau..89..261D,2017Icar..283...92W},
on Earth \citep{2012AcAau..73..250D}
or anywhere in the solar system \citep{2016JBIS...69...88G,2017JBIS...70..454G}. 

Without constraints on the purpose of the visit, the amount of data and the method to transmit it somewhere, these schemes can make no strong arguments about probe's sizes or locations. As an extreme example, a monolith under the surface of the moon \citep{Clarke1953} would only be able to communicate (realistically) once raised above the surface. Similarly, for a probe hiding in a lunar crater, about half the sky is never visible; it could not transmit data back to homebase if home is in such a direction. For example, if the probe came from the direction of the galactic center, a place on the moon's north pole would make (direct) communication impossible. A communication link analysis will bring up more such limitations (or preferences). Equally, it is strictly the additional assumption of data valuation which brings up the necessity (or strong preference) of nodes (relay stations) and certain physical constraints (location, size, energy level, mass, and wavelength) which can then be analyzed and optimized.

\subsection{Coordination schemes}
\label{sec:coord}
A set of schemes have been proposed for the scenario where the participants are usually located on planets, and are not aware of each other's existence. To determine the location and time for initiating a communication, numerous coordination schemes have been proposed, synchronized through 
the gravitational waves of binary neutron star mergers \citep{2019ApJ...875L..10S},
Gamma-Ray bursts \citep{1999PASP..111..881C},
maximum angular distance from the Sun \citep{2003AsBio...3..305C},
supernovae \citep{1976JBIS...29..469T,1977Icar...32..464M,1980Icar...41..178M,1994Ap&SS.214..209L},
pulsars \citep{2003IJAsB...2..231E,2017arXiv170403316V,2018IAUS..337..418V},
binary emphemerides \citep{1975Natur.254..400P},
and exoplanet transits \citep{1988ATsir1531...31F,2005ApJ...627..534A,2016MNRAS.459.1233K,2016AsBio..16..259H,2018MNRAS.473..345W,2019IJAsB..18..189F}.

All of these schemes have in common that the other ``side'' to connect to is located very far away. In contrast, the scheme presented here assumes a local entry point. As such, both ideas are complementary. Starting locally has multiple advantages: A possible return signal is on the order of days, not decades. Our technological capabilities are sufficient for in situ exploration within the solar system, but not beyond. If there is no local node to be found, more distant searches could be conducted afterwards.

\added{The idea of contacting nodes raises important issues related to METI (Messaging to ET Intelligence). There is an active debate about the question of whether Earth should initiate intentional transmissions to putative ET \citep{2014JBIS...67....5B}. In our case, it would open the question of whether Earth should try to contact putative nodes (or their locations) in our solar system, or in other systems. The more nearby a node is, the more drastic consequences could be expected. Taken to the extreme, METI also touches the question whether we should explore places like Cruithne with space probes, taking the risk of finding (and perhaps waking?) a dormant sentinel. 

There are convincing arguments that METI is unwise. The debate goes back to the fundamental question of who should do the broadcasting -- more advanced \citep{1974Natur.252..432B} and longer-lived \citep{1978Ap&SS..55....7B} species are argued to be better suited for this task. Yet, transmitting has an unknown (and perhaps zero) probability of success \citep{1984QJRAS..25..435W}. METI has been argued to be unwise, unscientific, potentially catastrophic, and unethical. For example, the motives of putative very advanced silicon AIs are unknown, and might be hostile \citep{2016JBIS...69...31G}. Even the reception of an alien message may pose an existential threat because decontamination is impossible \citep{2018arXiv180202180H}. On the other hand, it has been argued that ETI can already detect (some of) our leakage such as radio and radar transmissions. Earth can be observed from near and far with atmospheric studies and remote imaging, indicating the presence of intelligent beings on the planet. Finally, avoiding contact can also cause unknown risks, such as missing guidance that could enhance our own civilization's sustainability \citep{2016NatPh..12..890V}. Overall, METI policies are relevant even for solar system studies of nodes and probes.}

\subsection{Coloring a blank map}
Let us assume for a moment that the scheme presented here is correct: exploration probes are active around stars, and nodes exist in the gravitational lens planes to relay data. The big question then is: Where are the probes and nodes? Our current knowledge allows us to determine precisely the locations of stars (and some of their planets), but we have not yet found the network. Where do we search?

Imagine a blank map of the world. It contains the terrain, including rivers and mountains, but no borders and no cities. Where are the cities and highways on this map?

We might be able to guess some locations of major towns by making assumptions, such as the presence of gatherings near large streams, etc. In space, this would correspond to the assumption of stars with a rocky planet in the liquid water zone being of interest, and being visited by an exploration probe. These probes need to transmit and relay their data, and further guesses can be made about the major communication lines. For example, there might be a preference of interconnections between G-dwarfs over M-dwarfs due to larger habitability prospects. At the very least, we can make a priority list of network connections that appear logical to us, including stars like Alpha and Proxima Cen, Tau Ceti \citep{mccollum1992sails}, stars with known and interesting exoplanets such as the HabCat \citep{2003ApJS..145..181T,2003ApJS..149..423T}, the galactic black hole, etc. Would it make sense to locate relay nodes in every single system? This question highlights the need to understand how the network scales. How does the bandwidth increase with more and closer nodes? How much can an extra relay cost to make the marginal benefit of one more node worth it? This will be explored in a future paper of this series.

\subsection{Science drivers for probes}
The most obvious science driver for interstellar exploration probes is astrobiology. Another seemingly unexplored science case is astrometry with long baselines. Astrometric precision scales linearly with the aperture size of the collecting mirror, and with the baseline used for triangulation \citep[Eq. 4 in][]{2005ESASP.576...29L}. Gaia achieves $15\,\mu$as with $D=1.4\,$m and a 2\,au baseline. The distance to Alpha Cen is $276{,}000\,$au. Using even small interstellar baselines (1\,pc) with a meter-sized telescope gives a parallax precision of 0.1\,nas ($10^{-10}\,$arcsec). At a distance of 10\,Gpc, the apparent size of a central quasar engine is about 10\,nas \citep{2008PASP..120...38U}. Thus, interstellar astrometry can be used to measure parallactic distances to every visible object in the universe. Then, the exact structure and distances of galaxy filaments can be determined, shedding light on the big open questions in cosmology, such as dark matter, dark energy, and the Hubble tension \citep{2001ApJ...553...47F,2019ApJ...876...85R}.

\section{Conclusion and outlook}
This paper gave an outlook to a new series of papers on technical aspects of an interstellar communication network. The underling assumptions have been made explicit. While this summary is descriptive and partly speculative, subsequent papers will be very technical and focus on transmitters, repeaters, wavelengths, networks, and power levels. The overall purpose is to gain insight into the physical characteristics of an interstellar communication network to determine the most likely sizes and locations of nodes and probes. Feedback from the community is very welcome to shape and focus the next episodes.
\\
\\
\textit{Acknowledgments} I thank John Learned for asking about the technical properties of a galactic internet, which motivated part of this research; and Jason T. Wright for advice on how to split the topics involved into individual papers.
\bibliography{references}
\end{document}